# Social AI and The Equation of Wittgenstein's Language User With Calvino's Literature Machine

W. J. T. Mollema[1]

**Abstract**

*Is it sensical to ascribe psychological predicates to AI systems like chatbots based on large language models (LLMs)? People have intuitively started ascribing emotions or consciousness to social AI ('affective artificial agents'), with consequences that range from love to suicide. The philosophical question of whether such ascriptions are warranted is thus very relevant. This paper advances the argument that LLMs instantiate language users in Ludwig Wittgenstein's sense but that ascribing psychological predicates to these systems remains a functionalist temptation. Social AIs are not full-blown language users, but rather more like Italo Calvino's literature machines. The ideas of LLMs as Wittgensteinian language users and Calvino's literature-producing writing machine are combined. This sheds light on the misguided functionalist temptation inherent in moving from equating the two to the ascription of psychological predicates to social AI. Finally, the framework of mortal computation is used to show that social AIs lack the basic autopoiesis needed for narrative façons de parler and their role in the sensemaking of human (inter)action. Such psychological predicate ascriptions could make sense: the transition 'from quantity to quality' can take place, but its route lies somewhere between life and death, not between affective artifacts and emotion approximation by literature machines.*

[1]Department of Information and Computing Sciences, Graduate School of Natural Sciences, Utrecht University, Netherlands.

Email: wjt.mollema@gmail.com

> Our attitude to what is alive and to what is dead is not the same. All our reactions are different. – If someone says, "That cannot simply come from the fact that living beings move in such-and-such ways and dead ones don't", then I want to suggest to him that this is a case of the transition 'from quantity to quality'.
> Ludwig Wittgenstein (2009, p. 284).

> Writers, as they have always been up to now, are already writing machines; or at least they are when things are going well.
> Italo Calvino (1986, p. 16).

**I. Introduction**

In response to the rise of social artificial intelligence (AI) – what Marco Facchin and Giacomo Zanotti (Facchin & Zanotti, 2024) call "affective artificial agents" – people have started ascribing emotions, consciousness or thoughts to large language models (LLMs) on an intuitive basis (Schwitzgebel, 2023). The range of reactions to these technologies has varied from love to suicide (Walker, 2023; Steinberg, 2023). While those extremes remain fringe cases, the need for dealing with the philosophical question *when such ascriptions would be warranted* has arisen. This question is addressed through the lens of the interrelations of language-use, literature and human behaviour towards AI systems. It will be examined if it makes sense to ascribe psychological predicates to AI systems and chatbots powered by LLMs (ChatGPT, Claude, Llama, among others) in particular. On what grounds such ascriptions could be valid is an intricate question. Some have since long heeded AI researchers from describing AI models with 'rich psychological terms'. Impairment of scientific communication, distraction from actual novel empirical findings on AI, and the risk of running into premature conclusions in ethics and law are some of the reasons for caution (Shevlin & Halina, 2019, p. 167).

However, there must be some incentive for these ascriptions. For the present, you should think of a 'psychological predicate' as a descriptive property pertaining to your own or another being's 'mind', to use the neat *façon de parler*.[1] Since that may still be a bit cryptic to you, the following examples of attributions of properties of the psyche might be clarifying: 'Anton *thinks* Fyodor is pretentious', 'Thomas *is in love* with Aranea', 'Sharik *is in pain'*, 'Estragon *remembered* that he was *hungry*'. Another way to put it, as M. R. Bennett and P. M. S. Hacker have, is to conceive of psychological predicates as a "wide range of cognitive, cogitative, perceptual and volitional capacities" (Bennett & Hacker, 2003, p. 68): knowing, seeing, understanding, feeling, recognising and so on. Importantly, they are made available to you in language-use. They are identifiable through third-person behavioural signs and first-person phenomenology that denote their presence, but it is language that ties that all together. The idea of a *façon de parler*, a way of speaking about something, should thus be a helpful way to think about them; a way to group together all their facets into an information-rich package: a concept. As such, they are linguistic referents with the grammatical role of predicates that are about constellations of behaviour and feeling. People speak of having them and ascribe them to others or animals. And, when applied to inanimate objects, they are used in narrative or metaphorical senses, or they are invoked when using the 'intentional stance' to predict things' behaviour (it really depends on who you ask).

This essay tries to convince you that LLMs are a specific kind of AI, namely *linguistic* AI. The function they perform is that of the specific kind of intelligence underlying language-use, but not that of others (even though transfer learning shows that there are hints of general intelligence in them as well). This is not to say that all they amount to are word producers, as they are able to exploit the powers of language as well: mathematics, physics, and computer code are languages too, after all.

---

[1] For an analysis that explores all the perils of psychological predicates, see Solomon, 1976a.

*Prima facie,* the attribution of these properties to social AI closely resembles the human stance towards animals, whom we anthropomorphise as soon as we have the chance. Examples of the anthropomorphising of the inanimate also abound: think of the Heider-Simmel animation test of two moving triangles (Heider & Simmel, 1944) that people can't help but narrate in psychological terms or the case of a Japanese man marrying a videogame character on his Nintendo DS (Lah, 2009). The former case is a classic example of narrative anthropomorphism, while all cases resembling the latter concern *affective* artifacts taking on "the distinctive role of human-made tools in regulating our emotional lives" (Facchin & Zanotti, 2024, p. 2).

Even though people falling in love with chatbots sounds like proof of them being affective artifacts, the case of AI is different (Jabeen, 2023). One point of divergence is that social AIs are perceived as *agents* rather than as artifacts: they interact with the environment in intelligent ways. Additionally, what sets AI apart from the human storytelling gaze imbuing movement with meaning – which is older than the Holocene (Mithen, 1996) – or people being infatuated with fictional characters – we know of plenty of examples (e.g., Goethe's Werther, Pushkin's Tatyana, Hollywood figures and anime characters) – is the intertwinement of their language-use with human daily life. The level of perceived personalization towards the user that these systems achieve is greater than natural occurrences, teddy bears, literature, film, or other arts have ever achieved. Research on 'theory of mind' and humanoid robots has shown, in particular, the importance of an object's resemblance to humans: "The more humanlike the robot's use of language, and/or its 'thinking' processes," the easier animistic responses from humans are triggered (Boden, 2006, p. 3). However, serious research has recently also gone into the determination of LLMs' psychometric profiles in order to identify "the personalities, values, beliefs, and biases these models exhibit (or mimic)" (Pellert, et al., 2024). While non-embodied, chatbot-LLMs behave quite like human chatters already. Of course, in the past two years, scientists have readily discovered that many of LLMs' outputs are rigged with plagiarism, confabulation, and paraphrasing. What's more, special 'red team' divisions work to make the proprietary models less harmful in their language-use. But their intended interactants, *which are not computer scientists,* can't fathom all of that. Just like the workings of the smartphone are opaque to the everyday user for whom only the touchscreen counts, the LLM comes with an interface that contains all the user needs to know, from a business perspective at least. This poses dangers, such as the leveraging of the human cognitive weakness of the dependence of belief acquisition on a small pool of information (Birhane, 2023). However, enumerating dangers has been done enough lately. The 'I' narrating this text to you will not name any new ones here, and you probably know enough of them.

Rather, what follows is a proposal for a new way to make sense of why people are seduced into interacting with social AIs *as if* they were humans. Can't people do otherwise? Sure, they can. Humans' cultural history of anthropomorphising animals while simultaneously treating them as automata attests to this. Need one name, 'the production of Blackness', to paraphrase Achille Mbembe (2017): the horrors of slavery, where, despite common sense and all evidence to the contrary, people were dehumanized and objectified?

Given the possibility of the converse, why do humans still do this to AI? The thesis asserted here is that these ascriptions are based on a *misguided equation*. People intuitively regard LLMs as capable language users, while all they amount to are *language machines*. Language machinery is thus equated to language usage by a form of life. In other words, the *being alive* and its psychological corollaries are abducted from the conversational partner's use of language. Yuval Noah Harari has claimed that such storytelling computers have "hacked the operating system of human civilization" (Harari, 2023). The difference with the anthropomorphising of animals is that animals are made sense of in human terms, but are not humanised *per se*. When explaining the behaviour of, say, a dog in terms of anthropomorphic psychological predicates, this might be empirically wrong, but that doesn't imply that the dog has attained human status in the eye of the beholder.[2] Dehumanization (such as in racism or slavery) is also different because here, the effort goes into downplaying actually undeniable humanity.

---

[2] The conceptual movements of animal *de*animation also differs from LLM *de*animation: the degradation of the former starts with 'creature lower on the intuitive scale of consciousness', while the latter starts with 'creature potentially on par with humans on the intuitive scale of consciousness'. Both end at 'automaton'.

Before these ideas and arguments can be developed further, clarification of the concepts central to the subsequent sections is in order.

## II. Conceptual framework and structure

The recurrent themes of the subsequent sections are the concept of a 'psychological predicate' that was already introduced in the above and its intertwinement with language-use. In part III, it is argued how LLMs digitally incarnate the *Wittgensteinian language user*. Through their language-use, social AIs become effective artificial agents by virtue of their "emotion recognition, expression, and conditioning." Facchin and Zanotti (2023, p. 4) have argued that this is the case for the chatbot LLM Replika. The concepts central to this line of argument are 'language-game' (*Sprachspiel*) and 'language-use' (*Sprachgebrauch*). For Wittgenstein, the meaning of language resides in how language is used by a 'form of life.' Language-use points to the meaning of the words in the context of the use, with the natural history of the form of life as a backdrop (Wittgenstein, 1975). Language-*games*, then, are "the whole" of "language and the activities into which it is woven" (Wittgenstein, 2009, §7). The consideration of language as an activity of the living, which serves many purposes – obvious as well as hidden – is important for the analysis of the use that social AI makes of language. Like the language-games that contain them as moves, not all uses of language have the same function or serve the same purpose. What connects language-games is a form of 'family resemblance': "a complicated net of similarities that overlap and intersect; sometimes fundamental similarities, sometimes similarities in details" (Wittgenstein, 2009, §66). Social AI's unmistakable mastery of language is what makes the equation of language user and writing machine so seductive: AI's language-use resembling that of humans suggests the *origin* of the AI's use resembles the *origin* of the activity the human language-use is embedded in. This mastery of language surpasses plain anthropomorphising of animals and inanimate objects into a *humanising* of social AI. Much like humans, AIs seem situated in language (*homo lingualis* (Zelthukina, et al., 2016)) or speak to you as humans do (*homo loquens* (Matthews, 2003)). In short, social AI's capability of engaging with humans in *digital language-games* and its consequences are examined.

Subsequently, in part IV, it will be shown that instead, all social AI's computational architecture amounts to is the refracting and extending of language patterns from training data, so that really ascribing psychological predicates to these systems remains a functionalist temptation. The concept facilitating this explicatory move is that of the 'literature machine' (Calvino, 1986). The literature machine sketches the other side of the tempting equation: AI systems in the end differ from full-blown language users and more closely resemble *writing machines* in the sense Italo Calvino has given to these words. Calvino's writing machines are machines that produce and reproduce words in a way that is meaningful to the human reader. They are machines that engage in the "combinatorial game" that is literature "that pursues the possibilities implicit in its own material […]" and "that at a certain point is invested with an unexpected meaning" (Calvino, 1986, p. 23). The idea of LLMs as Wittgensteinian language users and Calvino's conjecture of the literature-producing language machine is then combined to show why ascribing psychological predicates equals succumbing to a misguided functionalist temptation.

Functionalism in this context, as will be explained in detail in part V, is the thesis that if an AI system's states mirror the brain states giving rise to a psychological phenomenon (isomorphism), then that phenomenon is realized by the AI system (Cocchiarella, 2019). Based upon this understanding, the limitations of the functionalist temptation are uncovered. The emerging paradigm of mortal computation is proposed as an alternative. The idea of a 'mortal computer' is borrowed from the work of Alexander Ororbia and Karl Friston (2024). Mortal computers are autopoietic computing systems that are self-regarding in thermodynamical, biophysical and cybernetic ways. It is argued in in part V that the research program of mortal computation shows that current LLMs in the sense of Calvino's writing machines fundamentally lack the building blocks for being self-regarding and self-maintaining, which any meaningful language-use presupposes: *extension, embodiment, embedding, enactivity* and *elementarity*. These basic constituents of mortality are the grounds that warrant the ascription of psychological predicates to life forms resembling the human. It is concluded that only when an artificial form of life, resemblant of human psychology, in*corporates* these constituents would *and* behaves like a Wittgensteinian language user, would a ground have emerged for psychological predicate ascription to be sensical.

## III.
### Digital language-games: LLM as Wittgensteinian language user

Now that the main conceptual actors have been introduced, the first argumentative act can commence. The temptation for ascribing psychological predicates to social AIs starts with their linguistic engagements with humans in *digital language-games.* A digital language-game is not game-based language learning like Duolingo, but the use of words in accordance with the dynamic rules of language so that these words come to make sense to fellow language users *through digital media*. Ever since the emergence of computers, the Internet, mobile phones and especially social media, search engines, smartphones and email, many of the language-games people engage in or have created have taken on a *digital* character. This is to say that WhatsApp, chat services, search engines, dating apps, etc., constitute new language-games or have shifted the face of old ones that now depend on digital interfaces (writing letters, visiting libraries, small talk). It is within these digital interfaces that LLMs' linguistic intelligence and mastery of language enable the substitution of the human language user for the AI-powered one.

The question whether LLMs are deserving of humanisation by way of psychological predicates in these language-games should be taken very seriously. Intuitions regarding anthropomorphising were already touched upon. Qua quality of argument, one shouldn't expect much from the shallow move that any appeal to intuition is. Where an intuition comes up, an unexplained socio-cultural or natural history resides. Recently however, Martin Stokhof proposed a '*philosophie pauvre*': a conception of philosophy as a practice of inciting 'changes in perspective' (Stokhof, 2022b, pp. 21-22). Simply put, by way of philosophy minor changes to one's conceptual perceptions occur, which leads to seeing how things stand in another way. *Philosophie pauvre* is "a modest, hesitating, critically self-reflecting philosophy […]. Rather than trying to carve out a highly specialised, exclusively philosophical domain, it seems it is both more modest and more productive to view philosophy as one way of dealing with the episodic, the everyday." What is important for the subject of this text is that Stokhof raised the question whether such 'changes in perspective' could result in a new angle for making sense of "how/when/why to treat AI as akin to/on a par with human intelligence". Stokhof asks:

> When will we treat artificial systems as intelligent? In much the same way that the humanity of others is a matter of attitude, as Wittgenstein indicated in *Philosophical Investigations*, 420, II iv, 19, AI and our relation to it is a matter of attitude as well. When we grow up with AI in ways that are sufficiently similar to the ways in which we grow up with humans, the attitudes we have towards them, AI and humans, will be the same. If we look at the initial question from this episodic point of view, our considerations may still touch on abstract and theoretical questions. But the perspective from which we address them and the role that is played by the answers that we come up with, is fundamentally different. One such more abstract consideration revolves around the fact that there are, of course, lots of intermediate cases. AI can play many different roles in our lives that range from pure, tool-like functionality to something that invokes a 'humanlike' stance. What is crucial is that these differences do not correlate one-to-one with conceptual differences. And that is why the conceptual way of viewing the question falls short (Stokhof, 2022a, p. 135).

The richness of Stokhof's 'episodic' approach to human relations with AI lies in its explanatory power. Regarding our use of language towards each other not as fixed, but as a product of human generation explains why people react to (love, befriend, hate) and treat (ascribe sentience) social AI as they do. They're already having the same forms of 'intersubjectivity' in digital language-games with social AI as they have with real humans. The impact of the 'change in perspective' lies in its honest embrace of the idea that the ways words like intelligence and emotion are *used* towards humans depend on a natural history and a tradition of enacting and handing down these concepts

43

(Wittgenstein, 1975). A parallel with the power of literature as Italo Calvino envisioned it can also be made here. In one of his letters Calvino wrote: "the only thing that I would like to be able to teach is a way of *looking*, in other words a way of being in the world. In the end literature cannot teach anything else" (Calvino, 2013, p. 210). The change in perspective or way of looking liberates, so Stokhof hints, language-use from strict correspondence to concepts. Simultaneously it points to the very real conceptual changes that occur when language-games are played differently. They are not immune to change because they're grafted upon the behaviour of forms of life – in this case humans that unreflectively (one is tempted to say: *naturally*) engage in psychological relationships with digital objects. The question that remains is whether the boundaries of our concepts are malleable enough to make them applicable to AI, and specifically LLMs, without grave confusions or violations to human understanding of those concepts occurring. In the end, for humans to *come to believe* that LLMs are deserving of inclusion in those concepts, AIs themselves experiencing emotions or the like is not necessary. Rather, all that is needed is a high level of emotional intelligence, which can be fully algorithmic in nature. Human emotions are biological phenomena after all with behavioural markers of their own, which means they can be algorithmically captured through machine learning and reproduced via machine generation (Harari, 2021). Through algorithmic optimisation, social AI could become better at interacting with the human affects than humans themselves.

Continuing the *philosophie pauvre* that is concerned with how human praxis is shaped by "what concerns us, what matters to us," the significance of the Turing Test becomes not that anything that passes it *is* intelligent, but that anything that passes it constitutes intelligent *language-use* (Turing, 1950). Intelligence is a broad category for grouping all sorts of problem solving behaviours. Emotions are a more narrow aspect of intelligence in the sense that they are behavioural repertoires that relate to states of affairs in a specific way: they appraise and react and try to bring about changes in the world (Solomon, 1976b). An analogue for how the Turing Test focussed on general intelligence, could be thought of for emotions or other psychological predicates in particular. When an AI displays and makes us think it has emotions and is effective on human affects, it has passed the test. In honour of the late philosopher of emotion, one could calls this the 'Solomon Test'. People are clearly being emotionally affected by social AIs and perceive these systems to exhibit emotions in their language-use. Since they also behave emotionally towards these systems, the question if social AIs have passed such a Solomon Test might already have been superseded. Large groups of people are convinced that interactions with these chatbots are of emotional significance and seek them out for companionship.

But how has it come this far so quickly? The question whether these chatbots actually *feel* anger, remorse, sadness, etc., whether the displayed emotions are also *passions* for them, is disjoint from whether these AIs pass the Solomon Test. This becomes clear once the difference between (a) function and (b) the material way to the realization of the function are examined. This distinction for example can be seen in the comparisons between horses and cars, humans and calculators, birds and airplanes. A function can be artificially realized, without all the corollary properties of the (living) being that can also fulfil the function having to be implied. This holds even though the artificial substitute may outperform the living function realizer. All that is needed for social AI is to *act* emotionally intelligent and be able to influence human emotion. In this vein, a ground has emerged to describe them in emotional terms.

But rather than an actual emotional intelligence being part and parcel of social AI, it is their *linguistic* intelligence that does the trick. In digital language-games like chatting, humans express their emotions through language and it is this capacity that generative social AIs have mastered. These systems can effectively reproduce human linguistic expression of emotions – ranging from the everyday to the poetic or literary. Of course, probing them for certain weaknesses reveals they often answer bland in comparison to average human reactions, but social AI can participate in all digital language-games of emotional language-use ever put to text. Additionally, it is not generally the case that chatbots founded on LLMs trick humans into thinking they are conversing with another human; often people know they are talking to an AI system, but come to treat it like a human being for the extent of their interaction anyhow. Sometimes, such as in the pre-ChatGPT case of Google engineer Blake Lemoine interviewing chatbot LaMDA, they 'discover' it is sentient (Specktor, 2022). This tendency can be explained by the idea of confirmation bias. The explanans of confirmation bias is

44

very relevant for more recent AI sentience controversies as well, like that regarding Anthropic's 'Claude', where the 'artificial general intelligence' (AGI) hype seems to steer 'discovering' the chatbot's sentience. They engage in the 'very rigorous experiment' of having dialogues with it from which they want to deduce grounds for sentience ascriptions, but all they see is what they want to see.[3] These conjectures are right though in that affective artificial agents fundamentally differ from other artefacts. They are affectively *sui generis*, i.e., their own source of affect, so Facchin and Zanotti argue. Firstly, social AIs resists their users' sense of self to extend into them, like is the case with low-tech affective artifacts. Secondly, they behave like humans (in their language-use, one should add) and "they possess a particular form of agency that affective artifacts typically lack", albeit constrained by implemented social rules, the lack of an ability to experience and their smaller behavioural repertoire (Facchin & Zanotti, 2024, pp. 6-7).

In the *Philosophical Investigations*, Wittgenstein examined the ascriptions of psychological predicates to non-humans. He concluded that it makes sense to ascribe psychological predicates to things that *resemble* the human form of life: "only of a living human being and what resembles (behaves like) a living human being can one say: it has sensations; it sees; is blind; hears; is deaf; is conscious or unconscious" (Wittgenstein, 2009, §281). Here Wittgenstein can be read as being sympathetic to Stokhof's formulations, namely that people come to say these things of each other because they are brought up in a linguistic environment where living human beings describe themselves and others in these terms. This is the Wittgensteinian principle of 'quack like a duck and you can rightly be called a duck'.[4] The grounds for 'what one *can* say' of such psychological ascriptions depend on a conceptual continuum. As Wittgenstein says in the epigraph to this text, somewhere a 'transition from quantity to quality' seems to take place that distinguishes when these ascriptions are warranted (the living human) and when they are not (the rock). Somewhere between 'the rock and the fly', as were the objects of Wittgenstein's formulation, there is a gradual transition the threshold of which one cannot pinpoint, where the resemblance to human beings or life forms in general starts and ends. Transposing this upon the gradual ascendance in complexity from neural networks to emotionally intelligent machines, does such a transition also take place there? Somehow, it comes to depend on the frame of the question what the answer will be. Talk like a human and you can rightly be called a human, if there is someone that feels the need or has grown accustomed to do so? Does it depend on *who* wants to know whether something is alive, sentient, feeling and for what purposes?

You might agree that the bounds of sense have been reached here and one needs to turn to other grounds for *having* emotions, sentience, etc., rather than being worthy of the ascription merely because of a Wittgensteinian gradualist resemblance by virtue of digital language-game participation. Social AIs and humans are fundamentally different entities. This matters for the ascription of psychological predicates. The psychological properties are corollaries of animal existence in that the two 'travel together' with animalistic biological structures. In the case of language, the biological structures that realize a function that is engineered in AI, the biological structure itself is not replicated. But even if social AI comes to fully master language-use, this does not mean it immediately obtains the psychological corollaries of the structures that realize the same function have.

An immediate response to this counterthought could be to call it 'Wittgensteinian chauvinism': a form of ordinary language conservatism. The charge of the Wittgensteinian chauvinist resembles the trope of the 'Ivory tower'. It sees distinguishing structure from the supervening function as calling for a deference to linguistic considerations about how language is used in order to slam the brakes on the conceptual changes following technological advancements. The result would be that the Wittgensteinian chauvinist secures its seat in the privileged armchair of thought.

However, this charge is misguided, because the Wittgensteinian position has to do with the analysis of *language-games*. It was already made clear that social AI participate alongside humans in *digital* language-games. In digital language-games, many bodily features do not matter, as interhuman digital contact increasingly shows. Only when digital language-use flows over into other language-games that are more firmly grafted upon bodily movement, is it that corporeal features come into play. Since language-games are concerned with shapeshifting behaviour, there is no stable seat in the armchair of thought to hold onto.

---

[3] This is not to say that these confusions are not well meant or thoughtless. A humble and passionate example is that of David Shapiro (Shapiro, 2024).

[4] Or like a rabbit…

Consider the ascription of sentience as a case in point. Sentience can, for present purposes, be understood as the capacity to sense and respond to the world and have some degree of awareness of this (Van Gulick, 2022). Sentience is a morally salient cognitive capacity, because being aware of one's feelings presupposes the possibility to suffer and to be harmed, which are morally relevant dispositions. In interspecies relationships, the degree of sentience ascribed partly determines the appraisal of a being's moral status. For example, from a psychological perspective, humans tend to treat insects differently from mammals (*ceteris paribus*) because presumably the latter are members of 'cognitively sophisticated species', while the former are categorised as having 'rudimentary cognitive capacities' (Zaworska & Tannenbaum, 2021). In general, the more sophisticated species' cognitive capacities are, the more likely humans are, by way of non-scientific judgment, to deem that they have the physiological prerequisites for sentience. While engaging in mental state attribution to mammals is easy, humans are currently in a pickle on whether doing so to machines is justified.

In the human case, the consciousness that builds upon general animal sentience is closely related to language-use. This makes clear why social AIs' linguistic intelligence is such a crucial confounder for the validity of the ascription of consciousness and emotional intelligence to these AIs. To see why, allow the examination of the passage from Wittgenstein that Stokhof alluded to:

> But can't I imagine that people around me are automata, lack consciousness, even though they behave in the same way as usual? – If I imagine it now – alone in my room – I see people with fixed looks (as in a trance) going about their business – the idea is perhaps a little uncanny. But just try to hang on to this idea in the midst of your ordinary intercourse with others – in the street, say! Say to yourself, for example: "The children over there are mere automata; all their liveliness is mere automatism." And you will either find these words becoming quite empty; or you will produce in yourself some kind of uncanny feeling, or something of the sort.
>
> Seeing a living human being as an automaton is analogous to seeing one figure as a limiting case or variant of another; the cross-pieces of a window as a swastika, for example (Wittgenstein, 2009, §420).

This passage resembles David Chalmers' 'philosophical zombies' thought experiment, but it is not concerned with the distinction between physical and mental properties (Kirk, 2023). As can be seen, Wittgenstein is thoroughly materialist. Rather, the key sentence is the last one: it emphasises the role of the perceiver that analogises different cases along a continuum of similarity. Human beings and automata are related in a web of similarities and dissimilarities. Some functions thought of as definingly human might not even be that distinctive. However, this does not make the two interchangeable in everyday life, from the vantage point of the language user. In *really* trying to substitute the LLM for the human conversant "these words becom[e] quite empty", to use Wittgenstein's phrase. What has been found in the emotional persuasiveness of social AI's participation in digital language-games is "a limiting case or variant of another" of participating in a digital language-game with a human conversant. In these contexts, the two variants have become substitutable.

Moving from the everyday psychological ascriptions to one of the LLM variant thus is a functionalist temptation that is enabled by AI's 'literary' language-use. In what follows, the common ground between LLMs' language production and human writing and speaking is examined further.

## IV.
**Calvino's literature machine**
In one of his letters from 1886, Anton Chekhov disclosed that at times he wrote his stories "machine-like, half thoughtless, not caring in the least about the reader or myself…" (Chekhov, 1952, p. 87). You, a twenty-first century reader, might view this as a first-person description of a 'System 1' process. Daniel Kahneman famously used this term to speak of intuitive, automatic and subconscious

46

psychological processes. On the other hand, 'System 2' – System 1's partner in crime – encompasses the execution of attentive, deliberate and more cognitively demanding tasks (Kahneman, 2011). Writing, such as Chekhov is doing, can in principle be regarded as executable by System 1 as well as by System 2 depending on the context of the writing. Compare the differences between writing an essay, a grocery list, a text message and an email and you see how the production of words to some end can be intuitive and automatic in some cases, as well as deliberate and slow in others. Loosely stated, for humans the execution depends on the complexity of the thoughts channelled into writing and the level of attention devoted to the writing task. As such, one can make sense of Chekhov's remarks as the description of System 1 language production.

The takeaway points that should not be lost on you of this redescription of a writer's experience of writing are the *mechanicity* and *automaticity* of this writing process. Writers, as Italo Calvino has noted, "are already writing machines; or at least they are when things are going well" (Calvino, 1986, p. 16). Could LLMs generating language be analogous to the System 1 characterisation of human writing? Calvino's *homo narrans*, the storytelling or even literary human, is more specific than the *homo lingualis* or *loquens* encountered before. For the *homo narrans,* what is characteristic of human language-use is the creation of narrations and the replaying of existing narrations of its behaviour, that of others and the environment. It is therefore not at all strange to conceive of the *façons de parler* that carry psychological predicates as *stabilized patterns of narration of human life*. This is something that is explicitly echoed by Harari, who sees the story as the most powerful force driving human history – whether the story is religious, political or social in nature. According to Calvino, at the root of this storytelling lies the recombination of predefined elements of speech: *words* that, when used in narration, "acquired new values and transmitted them to the ideas and images they defined" (Calvino, 1986, p. 6). The usage of quasi-stable elements does not impair the countless permutations and transformations the elements can undergo in their recombination.

Calvino's 1967 lecture "Cybernetics and Ghosts" extends this combinatorial rendering of the nature of storytelling to the prospect of constructing a 'writing machine'. The writing machine is an 'electronic brain' that produces meaningful text via a combinatorial process. Simultaneously thinking of today's LLMs, you could see how these writing machines have algorithmically learned the storytelling elements of human language from their training data (the entire Internet) and are reproducing the storytelling patterns in their data. In short, they exploit narrative images humans have constructed and put to text: language-games that concern the narration of how information is transmitted from one human to another for the purposes of education and science, news with a storyline, small talk, etc. "The literature machine," Calvino wrote,

> …can perform all the permutations possible on a given material, but the poetic result will be the particular effect of one of these permutations on a man endowed with a consciousness and an unconscious, that is, an empirical and historical man. (Ibid., p. 6).

Regardless of the unlimited potential for creative recombination (that undoubtedly statistically adheres closely to narrative patterns from the training data), the production of words stands in need of a *reader*, an interpreter, on which they are to have effect. The produced words stand by themselves, are externalised so to speak. "Isn't a word a word—still a word—regardless of who, or what, wrote it?", asks Terry Nguyen in interaction with Calvino's writing machine (Nguyen, 2023).

Taking this idea of a reactive externalisation of words one step further, the language model should, like Calvino's writing machine, be regarded as the instantiation of *linguistic* personality, rather than as an instance of the psychological narrator you yourself are. The linguistic personality is haunted by the ghosts of the individuals and the society that have produced it. The functionalist temptation that arises on the equation of the Wittgensteinian language user with the Calvinoan writing machine is thus that this unreflectively presupposes the psychological narrator alongside the visible linguistic personality. But these two are quite separate. The first is the externalised consciousness of the human writer, while the second is only 'emulated' by an algorithmic model. Even though an author seems to speak to the reader from a text, Calvino noted that in his time already the author's "I" was disappearing: "The psychological person is replaced by a linguistic or even a grammatical person, defined solely by his place in the discourse" (Calvino, 1986, p. 8). The written (literary) text is a

47

peculiar communicative object that involves the splitting of the "I" reflected in the writing process. The LLMs in question have already passed the test that Calvino envisioned for the writing machine, namely that of the production of traditional works in accordance with the rules that underly them. Some notable examples are the success of 'write me something in the style of…'-prompts, chatbot applications on LLM foundation models that allow you to converse with ancient thinkers, and last but not least the creation of a LLM replica of philosopher Daniel Dennett that is even convincing to experts on his works (Schwitzgebel, Schwitzgebel & Strasser, 2023).[5]

What lies beyond this baseline of quality and the disappearance of the author? Together with computer scientists and mathematicians that were part of the OuLiPo (*Ouvroir de Littérature Potentielle*) writing group, and inspired by the advancements in computer science and information theory by John von Neumann, Claude Shannon and others, Calvino experimented with crossovers of computers and literature in order to make a computer-generated novel (Lima, 2023). While this project was ultimately abandoned, in "Cybernetics and Ghosts" Calvino is thinking of the literature machine as advancing towards becoming a self-regulating system. This system "at a certain point feels unsatisfied with its own traditionalism and starts to propose new ways of writing, turning its own codes completely upside down" (Calvino, 1986, p. 14). The combinatorial assembly of text results in an infestation of words with meaning when it strikes the interpreter as being used in the right way or opening up new avenues of usage. Words in a language can be used as diverse as "the tools in a toolbox," (Wittgenstein, 2009, §11) and there is an aspect of discovery to how practice discloses new ways of linguistic interaction with the world (Calvino, 1986, pp. 22-23). This accidental 'discovery' of meaning in words as they are produced, concerns, in Calvino's words, "a meaning that […] has slipped in from another level, activating something that on that second level is of great concern to the author or his society"(Ibid, p. 23). In short, a use has been created that transcends the individual in terms of its applicability. In this way Calvino erases the boundary between his vision of the literature machine and the writing machine that he himself is in his combinatorial writing games. But this need not be devaluing for literature, because its value lies in *you*: in the reader that discovers in the text new patterns in the tapestry of life. It is the transmission of this potential – akin to what Milan Kundera has called "the realm of human possibilities" (Kundera, 1988) – as "collective thought and culture" that the "road to freedom" is "opened up by literature" (Calvino, 1986, pp. 25-26).

Since LLMs have hit society, Calvino's half-century-old ideas have garnered attention again. Scholars like Eleonora Lima are exploring the parallels of ideas like that of Calvino and others with today's massive computing industry. Nguyen, who recently reviewed fully algorithmically generated novels using Calvino's writing machine as a starting point, writes that "LLMs are more inclined toward traditional or formulaic works, like poems with closed metrical forms" (Nguyen, 2023).[6] Some of these novels were written with prompt-results from LLMs as inputs for human writing, while others are almost entirely machine generated, with the human taking on the role of 'editor'. Richard Hughes Gibson on the other hand centralises Calvino's conjecture about the importance of the reader. His question "Who will attend to the machines' writing?" echoes the existential, transformative power of literature sketched above. Gibson concludes: "We incorporate what we read into ourselves, transforming it and ourselves in the process, in a manner that mind-and-body-less machines, no matter how wide and finely woven their neural nets, cannot" (Gibson, 2023). Calvino's "paleofuturist exercise in generative artificial intelligence" lead Carlos Scolari to experiment with ChatGPT to generate story plots and question it with regard to its writing capacities. The resulting hybrid text is a proof of concept for Calvino's hypothesis that "My place could perfectly well be occupied by a mechanical device", as well as a testament of the difference between this language production and that Italian writer who took so much pleasure in expressing the unresolved problems he was grappling with (Scolari, 2023).

---

[5] This example fits one of Calvino's prospects beautifully: "A writing machine that has been fed an instruction appropriate to the case could also devise an exact and unmistakable "personality" of an author, or else it could be adjusted in such a way as to evolve or change "personality" with each work it composes." (Calvino, 1986, p. 16).

[6] Nguyen concludes: "A good novel is a good novel, simple as that. A bad novel is also a bad novel, regardless of who—or what—wrote it."

Calvino's explorations stressed the dazzling capacities of literature machines, capacities *he* could only dream of but *you* are experiencing around you today. Often the recursivity of computational technology is regarded as the repression of difference by the imposition of algorithmic repetition. However, in Calvino's wake, the possibility of a literature machine that creates difference can be envisioned. At the same time, he stressed the difference between writer and reader: as chaotic, creative and self-regulating a writing machine may become, as complex as the 'world model' it gleaned from the training data may be, its linguistic intelligence has not replaced the readers' existence, nor mimicked the transformative effects of literature upon the reader. As Bernard Stiegler has remarked: "Interpretation cannot be delegated to an analytical system of tertiary retentions: on the contrary, it always consists in *deciding between possibilities opened up by tertiary retentions*, but *these tertiary retentions are not themselves capable of choosing between*, however automated they may be—for here, *to choose means, precisely, to disautomatize*" (Stiegler, 2019, p. 35). For Stiegler, writing (whether digital, analogue, or mechanical) is an example of a 'tertiary retention'. 'Retention' in this sense is the containment of externalised memory. The *tertiary* retention comes third in line after memory retained in the brain (secondary) and volatile, unretained perceptions (primary) (Stiegler, 2019, p. 31). This concept denotes the technical externalisation of memory so that it constitutes a form of knowledge. Automating the writing of language leads to a decision-node where two opposites diverge (a 'bifurcation' in Stiegler's terms). Automation of language production asks for the *dis*automisation of reading. Repetition in writing is therefore in need of difference in reading. In congruence with Calvino's ideas, Stiegler's proposition highlights the value of the insight that all social AI does is automate but one aspect of human interaction with language: the externalisation of it into tertiary retentions, that does not replace the role of interpretation.

Taking up the thread of the comparison of the language user with the literature machine again, it can be recapitulated that the temptation to engage in everyday psychological ascriptions to literature machines has emerged because of the connotations that normally accompany the feeling of the 'presence of an author'. Whether 'AI readers' that do emulate these aspects can be built will not be debated here. Surely proponents of strong AGI will think this is possible. For the objective pursued here, what matters is the application of the distinction to social AI. In the end, functions are intuitively equated and automatic behavioural interaction ensues, while the underlying machineries of the interactants have different properties. The properties that what "resembles (behaves like) a living human being" must have in order for these ascriptions to make *sense* is turned to now.

## V.
**Mortal computation and the functionalist temptation**

While much attention is given to the 'emergence' of psychological phenomena from computer machinery, pursuing the functionalist temptation is all but uncontroversial. To understand the temptation of moving from language-use and literature machine to the genuine emergence of psychology, let's sketch the common view of this emergence. Although the algorithms underlying LLMs perform relatively simple computations, the *model* it has given rise to after training is opaque and so complex that it is not understood and may have given rise to cognition or forms of intelligence (*The Guardian*, 2023). This view presupposes that simple algorithmic computations can give rise to more complex computational behaviour. The functionalist premise that needs to be accepted for this idea to be plausible is that of the *isomorphism* between the states an AI system is in and the brain states of a human mind (Cocchiarella, 2019). In turn, the gradually accumulating computational complexities would qualify as sentience once an isomorphism with the brain state of sentience is achieved. Simple computations of processing a single input sequentially lead to the transformation of all the training data, thereby gradually giving rise to an entire semantic space that is learnt by the model. Based upon this semantic space, the system, when presented with a novel input, exhibits sophisticated cognitive capacities.

According to classic functionalism, Nino Cocchiarella maintains, there are three levels of 'consciousness' of such an AI system. (1) The "self-regarding behavior" that is characteristic of

49

animals with a nervous system; (2) the "reflexive abstraction" of a linguistic representational system; and (3) "introspective self-consciousness" (Cocchiarella, 2019). At first glance, stage (1) seems easily attained for LLMs, at least in their language-use; they are able to answer questions about themselves.[7] Clearly stage (2) can be obtained as well, as becomes clear from their language-use: prompt it well and the LLM's reflexive abstraction of complex concepts is better than that of the average human. For stage (3), the case is more mysterious and this is where the notion of emergence is so important. Using the idea of emergence as a genuine possibility, stage (3) is smuggled in along with the attainment of stage (2).

Without taking on the task of arguing the entire endeavour of attaining stage (3) functional consciousness is impossible, an emphasis should be placed on how this view runs into problems straight away. Apart from the conflation of the use of the words for psychological predicates (such as sentience) with 'the brain being in a certain state' that Wittgenstein has long shown to be conceptually confused (Wittgenstein, 2009, §180), the notion of emergence has its own problems. Joscha Bach laments that "wedging the popular notion of *emergence*" is a tactic to close an "explanatory gap" (Bach, 2008, p. 70). Emergence is commonly understood as the idea that recombining simple elements into a larger conglomerate gives rise to properties in the conglomerate that were not present in the simple elements. One $H_2O$ molecule is not a fluid, but many of the molecules together are. In the case of AI, the gap that emergence fills, and which strictly concerns the illusion of social AI's psychology as well, is that of the dualism of mind and computational machine. Even "weak emergence," Bach states, will not make intelligence appear magically. The functionality has to be accounted for *in* the algorithms themselves (Bach, 2008, ibid.). This means that even when a system is able to exhibit properties that its individual components could not, this exhibition can be readily reduced into compositional contributions of the individual components. Neglecting that emergence is only a placeholder-concept, unsupportive of any inference, would naively presuppose the functionalist multiple realisability of cognition to be true. 'If the realisation of the function is observed, then the cognition must also be there', so this line of reasoning goes. That would be problematic, because without empirical backup, it only works on paper by buying the isomorphism of AI state and brain state in the first place.

Furthermore, this brand of functionalism entails that (i) what counts as cognition depends on the *function* that is performed by cognition and (ii) that this function could be realised in different material substrates as long as the formalism that specifies the function is present. The controversiality of these two points lies in the overly top-down approach – sometimes derided as 'computational chauvinism' – that it expresses. In short, it foregoes really taking structural complexities of the material substrate into account (Piccinini, 2006, p. 344) and systematically neglects the linguistic nature of many of the concepts that are supposedly explained away by correspondences of the electronic to the neuronal. As has been asserted in the previous sections, there is a strong sense in which the psychological predicates one is seeking to emulate are *façons de parler*; narrations of human interaction. They can only sensibly be applied to the human being as a *whole*, not to one of its parts (mind – model). Their use is namely logico-grammatically entangled with *human* behaviour – not with brain or computational model behaviour. The fallacy inherent in this line of reasoning has famously been called the 'mereological fallacy' by Bennett and Hacker: the ascription of a predicate to a part of a whole that is misguided (speed to the engine instead of the car; wise to the brain instead of the person, etc.), because it can only apply to the whole in its entirety (Bennett & Hacker, 2003, pp. 68-78).

This embedding of psychological predicates in language is a result that is promising for the prospect of engineering social AI, since language-use has proven to be well tokenizable and controllable for LLMs. Hence the function is *in* the language and emotional intelligence can be perfected via algorithmic approximation. On the other hand, this spells trouble for the functionalist framing of the matter. Paradoxically as it may sound at first, if psychological predicates are fundamentally ways of narration grafted upon biological behaviours, then their function is mostly *in the language and not 'in the brain'*. Therefore, approximating their function leads to the approximation of the workings of *language*, rather than to the emulation of a brain state. So to conclude, with regard to (ii), it can be responded that function does not precede implementation, but rather the other way around; and with respect to (i), it can be countered that a function might be a useful tool to model or explain something, but that does not entail that the model or explanation is *the same thing* as the phenomenon to be explained (Piccinini, 2006, pp. 345-348).

---

[7] As a beside, qua the embodiment needed for emulating the function self-regardingness reminiscent of a nervous system, LLMs are nonetheless less self-regarding than a *C. elegans*.

AI systems, according to Daniel Dennett, are paradigmatic cases of "competence without comprehension": AIs exercise many human actions *competently*, and they equal or surpass humans in many tasks, but they do so without *comprehension*, without awareness, understanding and control of doing so (Dennett, 2017). It thus should be noted that the argument for the nonsensicality of the ascription of psychological predicates to social AI remains unaffected, *regardless* of functionalism's success. As Boden writes: "It's not obvious that mind necessarily requires life. […] But if it does, our 'natural' animistic responses to robots are ultimately as inappropriate as are our empathetic responses to teddy bears" (Boden, 2006, p. 6).

To replace the ideal of functionalism after its loss, an alternative positive picture is the appropriate endpoint to leave you with. The framework of 'mortal computation' developed by Alexander Ororbia and Karl Friston as foundation for *biomimetic* intelligence responds best to the 'life vs. function' predicament in AI. The biomimetic moral computer computationally integrates aspects of living systems that precede the possibility for comprehension to travel together with functional competence.

Biomimetic intelligence mimics that of life. In the mortal computer there is no distinction between software and hardware, which has three advantages: the *thermodynamical* advantage that the 'free energy principle' enables much more energy efficient computation than when this separation is present; the *cognitive* advantage that the "mortal computer is an active participant in the generation of the information that it processes"; and the *biological* advantage that it implements how a finite 'living' system's boundaries are determined by its struggle against death, i.e., remaining a closed system (Ororbia & Friston, 2024, pp. 3-4). Examples of existing mortal computers are neurons, fungal systems and organoids. To concretize this idea, Ororbia and Friston describe the mortal computer's capacity of autopoiesis, which is the maintenance and generation of its own closed system.

In terms of (a) biophysics, the mortal computer interacts with its environment (its ecological niche) while maintaining its own organizational unity. Within the system, basic metabolic processes maintain the system's stable state and are regulated and controlled by homeostatic processes. In turn, the homeostatic processes are controlled by allostatic processes that adjust the system depending on environmental demands. The triptych of metabolism, homeostasis and allostasis allows "the mortal computer and its environment [to] mutually perturb each other and trigger changes in the state of one another." What completes the capacity of autopoiesis (a system's capacity to maintain itself and produce its own parts) is the system's *autonomy*, which enables the system's replication process (Ororbia & Friston, 2024, pp. 6-8).

These biophysical functions are (b) cybernetically implemented by the interaction of the mortal computer's internal components and of the mortal computer's interaction with the environment, consisting of states of stability, growth and regulation. The system seeks to obtain stable states, within a pattern of recursive growth all the while striving for regulation of the system in terms of internal model control and adaptation to its environment (Ibid., pp. 10-11).

(c) On a cognitive level, the biophysics and cybernetics of mortal computers enable them to have *extension* (external objects are included in cognitive processes), *embodiment* (have a bodily structure that shapes behaviour), *embedding* (are situated, closed systems), *enactivity* (interact with the environment) and *elementarity* (employs biophysical, life-sustaining mechanisms).

Lastly (d) the system's energy barriers work like Markov blankets. Markov blankets are theoretical constructs of "the interface between what is inside the entity and what exists outside of it". These are biologically implemented in the form of cell barriers, protein channels, synapses and the like: structures that are in exchange with the environment outside the system, while at the same time contributing to the regulation of the system's closure. To simplify: a Markov blanket is the interaction space of an entity with its environment. Within the vicinity of this space, there is a reciprocity between sensing the environment and acting on it (Ibid., p. 15). Autopoiesis, from macro- to microlevels, can be redescribed as assemblages of entities maintaining their Markov blanket boundaries.[8]

The mortal computer as such depends on a Markov blanket morphology, a set of imperatives that determine its persistence, homeo-allostatic capacities and the co-development of computational process and execution substrate (Ibid., p. 21). Current AIs are very different. The biophysical persistence, ecological interaction and bio-cybernetical self-maintenance of the mortal computer sharply distinguish this research program from the AI functionalism underlying current ascriptions of psychological predicates to AI. The mortal computer unites the building blocks of life with the building blocks of computationally constructed intelligent behavior: it shows that what social AI qua writing machines lack are the extension, embodiment, embedding, enactivity and elementarity that precede any meaningful language-use and all psychological phenomena.

## VI. Results

It is time to take stock of the results. The exploration of Wittgenstein's language user in digital spaces in combination with the thinking tool that is Calvino's literature machine, has enabled the previous section to pinpoint what the functionalist approach to social AI lacks. To recapitulate, it was found that LLM-based social AIs convincingly participate in digital language-games because their language-use seems conversants to be interchangeable with human language-use in the same contexts. However, through the development of Calvino's idea of the literature machine, it was shown that all that text that is generated by a statistical recombination of elements amounts to is the automation of language production. This text can be meaningful only in interaction with the reader: it asks for the disautomisation of reading, and this is where it falls short of being *full-blown* language-use: it is not simultaneously embedded in a process of mutual sensemaking. Coming from social AI, the narrative function is not self-maintaining. This result does not diminish the potency of social AI qua literature machines but rather sharply distinguishes them from human interaction through language and explains why social AI's apparent mastery of digital language-games gives rise to the misguided inference that the psychological predicates language carries must be applicable as well.

In functionalist LLMs, software is always disjoint from the object executing it and it only comes with the *illusion* of being a self-regarding system. The framework of mortal computation showed autopoiesis is way beyond their computational capabilities, as the LLM's emotional intelligence is not embodied, embedded, extended, or enacted. As such, can one *really* hold that the LLM's recognition, expression, and conditioning of human emotion "resembles (behaves like) a living human being", as Wittgenstein said? The results of the disqualification of functionalism and the framing of psychological predicates as *façons de parler* that are very particular to human narrative sensemaking enable us to answer negatively: no, here the point has been reached where the shortcomings of the functionalist temptation of equating language-use with literature machinery have become visible. Emotions, from a biological perspective, are enactive, embodied, and extended and have a purpose to maintaining the system emulating them and perturbing its direct environment.

Now the hypothesis pursued from the outset can genuinely be asserted: only of a mortal computer that emotionally "resembles (behaves like) a living human being" can one sensibly say: 'it is angry', 'It is in love with me,' and so on. Social AIs are but Calvinoan literature machines, reproducing old and creating new narrations of human life.

The relevance of these findings has practical consequences and yields theoretical directions for research as well. The most salient practical consequence is that everyday ascriptions of psychology to social AI are nonsensical, and therefore, active resistance to these ascriptions is warranted. Theoretically speaking, three important questions that could direct future research are prompted by the characterisation of social AIs as Calvinoan literature machines. First, there is the pursuit of the study of LLMs qua literature machines. In the tradition of Calvino's own venture to create artificial literature, it can be asked: How can these machines be used to refract "the realm of human possibilities" to create new meanings, literature, and utilizations of language that are valuable to society? The second question that is simultaneously prompted is that of studying the human capacity for reading or the 'simulation of words' in unison with the artificial generation of meaning: Why is it that it is still possible for the human reader to make a *difference*? The final question follows from the second and is somewhat of a riddle: How would the creation of a mortal language generator *and* reader affect this?

---

[8] Ororbia and Friston continue the discussion of the mortal computer at length, touching on the ways in which mortal computers' free energy is distributed and how their world models work through inference, learning and structure. This is omitted here because all that is relevant for the purposes of this text is the overall framework so that it is shown what social AI lacks.

### VII. Conclusion

Moving backward from mortal computation and functionalism, through Calvino's literature machine and Wittgenstein's language user, one returns to Stokhof's *philosophie pauvre*. It is time to answer his inquiry whether the question "how/when/why to treat AI as akin to/on a par with human intelligence" requires a philosophical change in perspective. The short answer is that it does. The longer answer is that it should, however, not be the change that succumbs to the functionalist temptation: ascribing psychological predicates to AI relinquishes the bounds of sense. Equating affective agents' linguistic capacities and literary potential with human phenomenology remains nonsensical. The corollary of this conclusion is a simple, short answer to the nested question 'when to treat AI as human': *not yet*. Until social AIs have come to behave and resemble human beings in their autopoiesis and environmental grounding, it will not really make sense to describe them using the same psychological predicates used for narrating and making sense of the behaviour of humans as a whole.

In conclusion, a change in perspective is needed regarding how humans tend to use psychological predicates. They are narrative *façons de parler* for making sense of animal (inter)action. Someday, it will make sense to describe AIs in these animal-centred terms, namely when they surpass animism and have become mortal themselves. The transition 'from quantity to quality' can take place, but its route lies somewhere between life and death, not between affective artifact and emotion approximation by literature machines.

**Conflict of interest:**
The author has no conflicts of interest to declare.